\documentclass[a4paper]{jpconf}
\usepackage{graphicx}
\usepackage{cite}
\def\gsim{\,$\raise0.3ex\hbox{$>$}\llap{\lower0.8ex\hbox{$\sim$}}$\,}
\def\lsim{\,$\raise0.3ex\hbox{$<$}\llap{\lower0.8ex\hbox{$\sim$}}$\,}

\begin{document}


\title{Ground-state phase diagram of an anisotropic S=1/2 ladder
with alternating rung interactions}

\author{
Takashi~Tonegawa$^{1,2}$,
Kiyomi~Okamoto$^{3}$,
Toshiya~Hikihara$^{4}$ and
T{\^o}ru~Sakai$^{5,6}$
}

\address{
$^{1}$Professor Emeritus, Kobe University, Kobe 657-8501, Japan\\
$^{2}$Department of Physical Science, Osaka Prefecture University,
Sakai, 599-8531, Japan\\
$^{3}$College of Engineering, Shibaura Institute of Technology,
Saitama, 337-8570, Japan\\
$^{4}$Faculty of Science and Technology, Gunma University, Kiryu,
376-8515, Japan\\
$^{5}$Graduate School of Material Science, University of Hyogo, Hyogo
678-1297, Japan\\
$^{6}$Japan Atomic Energy Agency, SPring-8, Hyogo 679-5148, Japan
}

\ead{tone0115@vivid.ocn.ne.jp}

\begin{abstract}
Employing mainly numerical methods, we explore the ground-state phase diagram
of an anisotropic \hbox{$S\!=\!1/2$} ladder, in which leg interactions are
uniform and isotropic, while rung interactions are alternating and have a
common Ising-type anisotropy.  We determine the phase diagram in the case where
\hbox{$J_{\rm leg}\!=\!0.2$} (antiferromagnetic),
\hbox{$J_{\rm rung}\!=\!-1.0$} (ferromagnetic) and 
\hbox{$|J_{\rm rung}'|\!\leq\!1.0$}, the first one being the magnitude of the
leg interaction and the second and third ones those of the rung interactions,
which are alternating.  It is emphasized that the system has a frustration
when $J_{\rm rung}'$ is positive.   We find that, in the frustrated region,
the Haldane state appears as the ground state even when the Ising character
of rung interactions is strong.  This appearance of the Haldane phase is
contrary to the ordinary situation, and it is called the inversion phenomenon
concerning the interaction anisotropy.  We also find that an incommensurate
state becomes the ground state in a portion of the Haldane phase region.
\end{abstract}


\section{Introduction}

The frustration effect on the ground-state properties of low-dimensional
quantum spin systems with competing interactions has long been a subject
of active research.  According to a significant amount of theoretical and
experimental effort which has been devoted so far, it is now widely known that
an interplay between two phenomena of great interest, frustration and
quantum fluctuation, leads to various exotic ground states.  A typical and
long-established example of these ground states is the dimer state
accompanying spontaneous translational symmetry breaking in an
\hbox{$S\!=\!1/2$} zigzag chain in which antiferromagnetic nearest-neighbor
(nn) and next-nearest-neighbor (nnn) interactions are competing with each
other~\cite{majumdar-ghosh,tonegawa-harada,okamoto-nomura}.

The effect of the frustration on the ground-state properties of an
\hbox{$S\!=\!1/2$} two-leg ladder system has been extensively studied in the
cases where additional leg nnn and/or diagonal interactions are competing with
the leg nn and rung interactions~\cite{frustrated-leg-ladder1,
frustrated-leg-ladder2}.  In the present paper, as another example of the
frustrated \hbox{$S\!=\!1/2$} two-leg ladder systems, we discuss the case where
the rung interactions are alternating~\cite{japaridze-pogosyan,amiri-etal} and
aim at exploring its ground-state phase diagram.  We express the Hamiltonian
describing this system in the following form:
\begin{eqnarray}
{\cal H}
  \!\!\!&=&\!\!\! J_{\rm leg} \sum\nolimits_{j=1}^{L}
              \bigl\{{\vec S}_{j,a}\cdot {\vec S}_{j+1,a}
            + {\vec S}_{j,b}\cdot {\vec S}_{j+1,b}\bigr\} \nonumber\\
      && +\; J_{\rm rung} \sum\nolimits_{j=1}^{L/2}
            \bigl\{\gamma(S_{2j-1,a}^x S_{2j-1,b}^x 
                          + S_{2j-1,a}^y S_{2j-1,b}^y)
                 + S_{2j-1,a}^z S_{2j-1,b}^z \bigr\} \label{eq:hamiltonian} \\
      && +\; J_{\rm rung}' \sum\nolimits_{j=1}^{L/2}
            \bigl\{\gamma(S_{2j,a}^x S_{2j,b}^x 
                          + S_{2j,a}^y S_{2j,b}^y)
                  + S_{2j,a}^z S_{2j,b}^z \bigr\}\,,  \nonumber
\end{eqnarray}
where ${\vec S}_{j,l}\!=\!\bigl(S_{j,l}^x,\,S_{j,l}^y,\,
S_{j,l}^z\bigr)$ is the \hbox{$S\!=\!1/2$} operator acting at the
($j$,$\,l$) site assigned by rung $j$ and leg $l(=\!a~{\rm or}~b)$;
$J_{\rm leg}$ denotes the magnitude of the isotropic leg interaction;
$J_{\rm rung}$ and $J_{\rm rung}'$ denote those of the two kinds of anisotropic
rung interactions which are alternating, the $XXZ$-type anisotropy being
controlled by the parameter $\gamma$ in common with both interactions; $L$ is
the total number of spins in each leg, which is assumed to be even.  It is
emphasized that this system has a frustration when
\hbox{$J_{\rm rung} J_{\rm rung}'\!<\!0$} irrespective of the sign of
$J_{\rm leg}$.

In the following discussions, we confine ourselves to the case
where $J_{\rm rung}$ is ferromagnetic, and we put
\hbox{$J_{\rm rung}\!=\!-1$}, choosing $|J_{\rm rung}|$ as the unit of
energy.  We also consider, for simplicity, only the case
where \hbox{$J_{\rm leg}\!=\!0.2$}, \hbox{$|J_{\rm rung}'|\!\leq\!1$} and
\hbox{$0\!\leq\!\gamma\!<\!1$}, that is, we assume that $J_{\rm leg}$ is
antiferromagnetic and relatively weak, and that the anisotropy of the rung
interactions is of the Ising-type.  Determining the ground-state phase diagram
on the $\gamma$ versus $J_{\rm rung}'$ plane, we mainly employ the
numerical methods such as the exact-diagonalization (ED) method and the
density-matrix renormalization-group (DMRG)
method~\cite{dmrg-white-1,dmrg-white-2} with the help of physical
considerations as well as already-known results for some special cases.

\section{Special cases}

We discuss here some special cases for which the ground states have already
been clarified or for which they are reasonably anticipated by physical
considerations.

\subsection{Case where \hbox{$J_{\rm rung}\!=\!J_{\rm rung}'\,(=\!-1)$}}

In this case where
\hbox{$0\!<\!J_{\rm leg}\!\ll\!|J_{\rm rung}|\!=\!|J_{\rm rung}'|$}, by the
use of the degenerate perturbation theory, the present
system can be mapped onto the \hbox{$S\!=\!1$} chain in the following way.
The eigenstates of an isolated $J_{\rm rung}$- or $J_{\rm rung}'$-rung are
given by $\phi_j^{(1,+)}\!=\!\alpha_{j,a}\alpha_{j,b}$,
$\phi_j^{(1,0)}\!=\!(\alpha_{j,a}\beta_{j,b}\!+\!\beta_{j,a}\alpha_{j,b})/
{\sqrt{2}}$,
$\phi_j^{(1,-)}\!=\!\beta_{j,a}\beta_{j,b}$ and
$\phi_j^{(0,0)}\!=\!(\alpha_{j,a}\beta_{j,b}\!-\!\beta_{j,a}\alpha_{j,b})/
{\sqrt{2}}$, where $\alpha_{j,l}$ denotes the \hbox{$S_{j,l}^z\!=\!+1/2$}
state and $\beta_{j,l}$ the \hbox{$S_{j,l}^z\!=\!-1/2$} state.  The
corresponding energies are, respectively,
$E^{(1,+)}\!=\!-1/4$, $E^{(1,0)}\!=\!(1\!-\!2\gamma)/4$,
$E^{(1,-)}\!=\!-1/4$ and $E^{(0,0)}\!=\!(1\!+\!2\gamma)/4$, for all
$j$'s.  Then, it is easy to see that when $\gamma$ is sufficiently large, the
state $\phi_j^{(0,0)}$ can be neglected.  We introduce the pseudo
\hbox{$S\!=\!1$} operator ${\vec T}_j$ for rung $j$, and make the
$T_j^z\!=\!+1$, $0$ and $-1$ states correspond to $\phi_j^{(1,+)}$,
$\phi_j^{(1,0)}$ and $\phi_j^{(1,-)}$, respectively.  The relation
${\vec T}_j\!=\!{\vec S}_{j,a}\!+\!{\vec S}_{j,b}$ holds, as is readily shown
by comparing the matrix elements of both operators
${\vec T}_j$ and ${\vec S}_{j,l}$ with respect to
$\phi_j^{(1,+)}$, $\phi_j^{(1,0)}$ and $\phi_j^{(1,-)}$.  Thus, the
Hamiltonian~(\ref{eq:hamiltonian}) for the \hbox{$S\!=\!1/2$} operator
${\vec S}_{j,l}$ can be mapped onto the effective Hamiltonian
${\cal H}_{{\rm eff}}$ for the \hbox{$S\!=\!1$} operator ${\vec T}_j$, which
is given by
\begin{equation}
    {\cal H}_{{\rm eff}}
         = 0.1 \sum\nolimits_{j=1}^L {\vec T}_{j} \cdot{\vec T}_{j+1}
                   + D \sum\nolimits_{j=1}^L (T_j^z)^2 \,,~~~~~~
       D = (\gamma-1)/2\,,
   \label{eq:effective}
\end{equation}
where the on-site anisotropy ($D$-) term comes from the difference between
$E^{(1,+)}=E^{(1,-)}$ and $E^{(1,0)}$.

It has already been clarified by Chen {\it et} {\it al}~\cite{chen-etal}
that in the \hbox{$S\!=\!1$} chain governed by the effective
Hamiltonian~(\ref{eq:effective}), the
phase transition from the N{\'e}el to the Haldane state takes place at
\hbox{$D\!\sim\!-0.04$} as $D$ increases. This suggests that in
the present \hbox{$S\!=\!1/2$} ladder, the ground state is the
antiferromagnetic stripe N{\'e}el (AFstN) state, sketched in the left of
figure~\ref{fig:phases}, or the Haldane state, sketched in the middle of
figure~\ref{fig:phases}, depending upon whether
\hbox{$\gamma\!<\!\gamma_{\rm c}^{{\rm (AFstN,H)}}$} or
\hbox{$\gamma_{\rm c}^{{\rm (AFstN,H)}}\!<\!\gamma\!<\!1$} with
\hbox{$\gamma_{\rm c}^{{\rm (AFstN,H)}}\!\sim\!0.9$}.  (It is noted that the
N{\'e}el state in the \hbox{$S\!=\!1$} chain corresponds to the AFstN state in
the \hbox{$S\!=\!1/2$} ladder.)

\subsection{Case where \hbox{$|J_{\rm rung}|(=\!1)\!\gg\!J_{\rm rung}'\!>\!0$}}

In this case, the procedure discussed in the previous subsection can be applied
to the $J_{\rm rung}$-rung only.  Thus, the Hamiltonian~(\ref{eq:hamiltonian})
can be mapped onto an anisotropic ^^ \hbox{$S\!=\!1$}'-^^ \hbox{$S\!=\!1/2$}'
diamond chain.  Hida and Takano~\cite{hida-takano} have studied the
ground state of this chain.  By using their results, we may conclude that, when
\hbox{$\gamma\!\to\!1.0$}, the ground state in this case is the Haldane state
if \hbox{$J_{\rm rung}'\lsim 0.26$}.

\subsection{Case where \hbox{$J_{\rm rung}'\!\sim\!-J_{\rm rung}(=\!1)$}} 

\begin{figure}[t]
  \scalebox{0.24}{\includegraphics{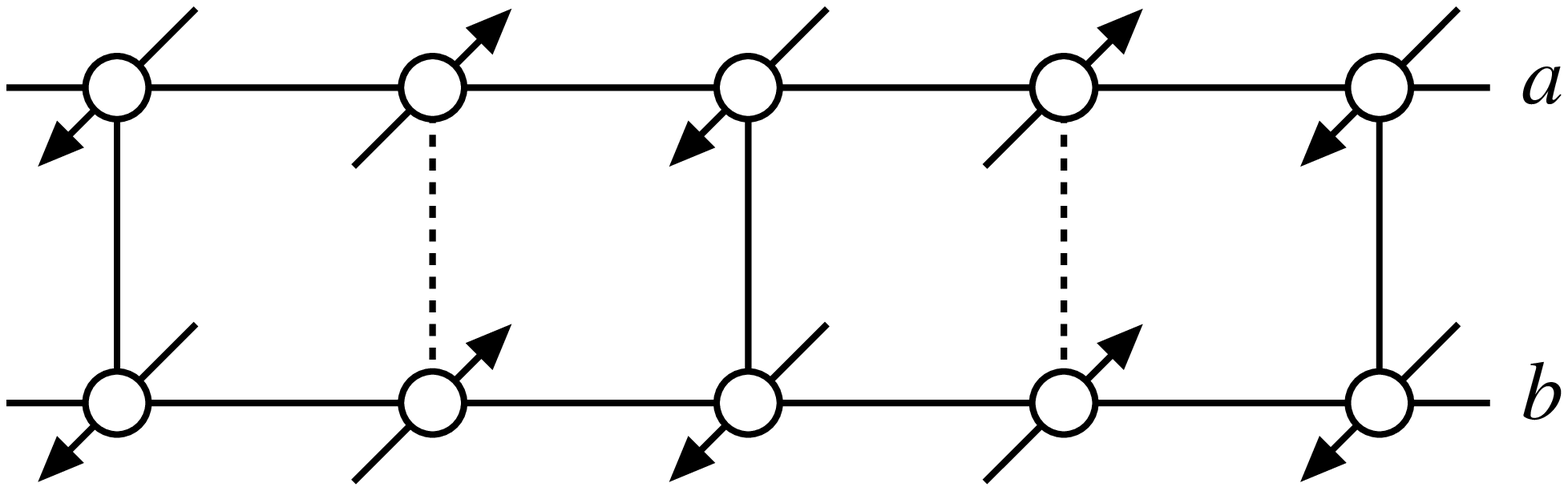}}~~~~~
  \scalebox{0.24}{\includegraphics{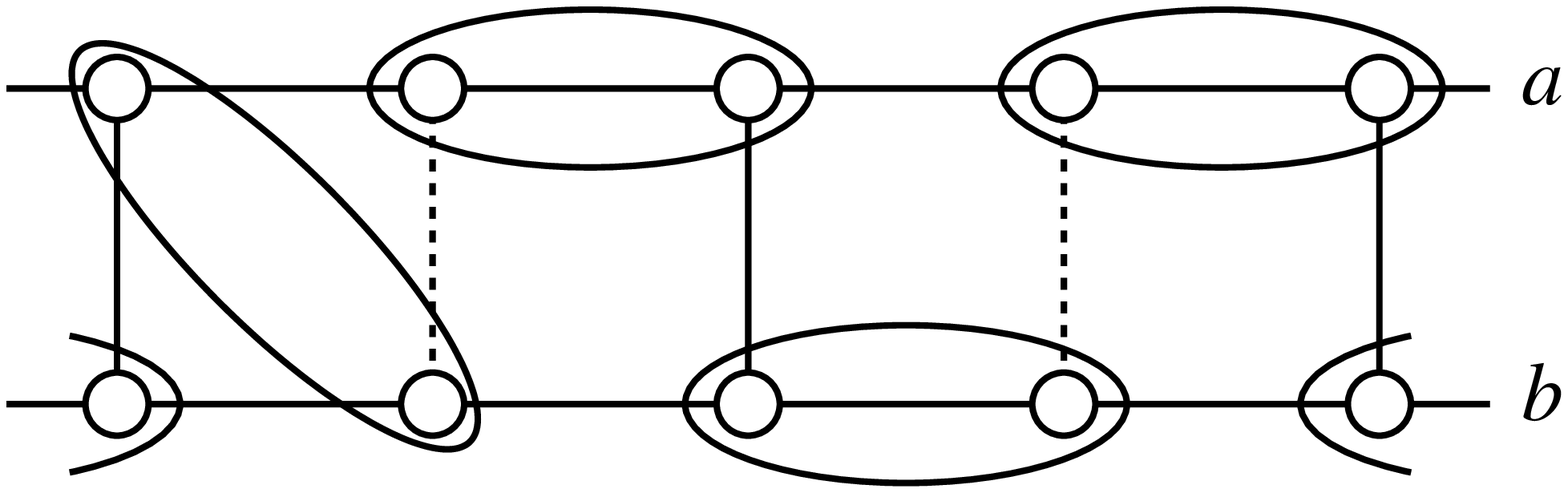}}~~~~~
  \scalebox{0.24}{\includegraphics{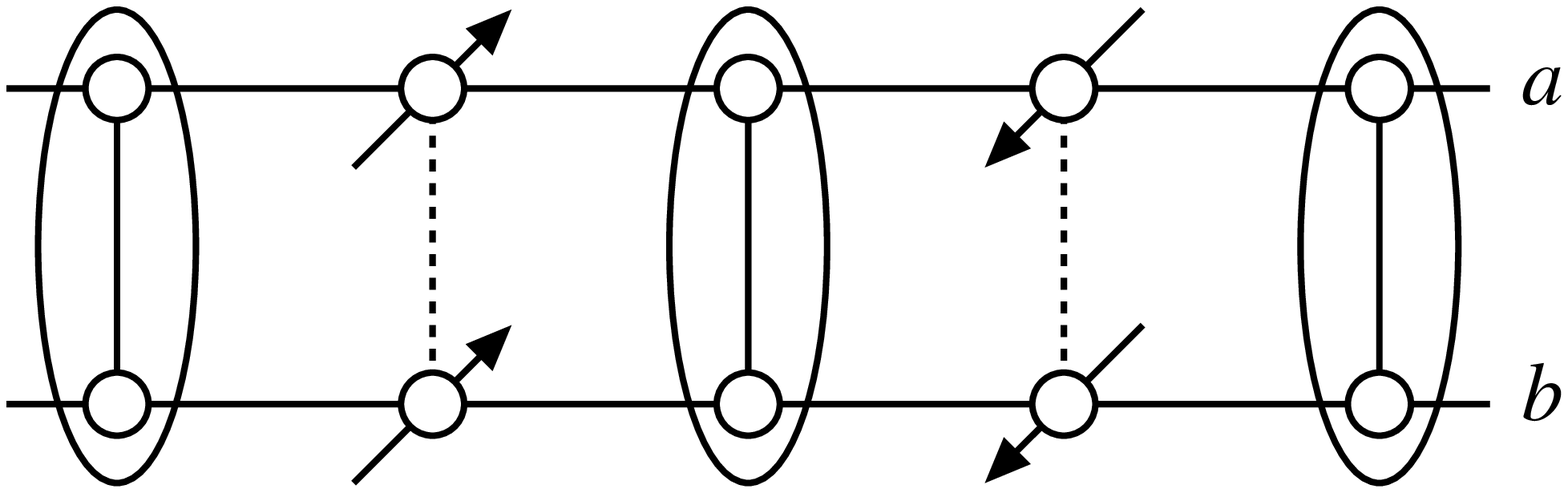}}
  \caption{Schematic pictures of the AFstN (left), Haldane (middle) and
           F-SD (right) phases.  Each of the small open circles denotes an
           \hbox{$S\!=\!1/2$} spin.  Two \hbox{$S\!=\!1/2$} spins in each
           ellipse form a singlet dimer.  An \hbox{$S\!=\!1/2$} spin with an
           upward arrow and that with a downward arrow are, respectively, in
           the \hbox{$S^z\!=\!+1/2$} ($\alpha$) and \hbox{$S^z\!=\!-1/2$}
           ($\beta$) states.
  }
\label{fig:phases}
\end{figure}

In this case where
\hbox{$0\!<\!J_{\rm leg}\!\ll\!|J_{\rm rung}|\!\sim\!|J_{\rm rung}'|$},
the lowest-energy state of an isolated $J_{\rm rung}'$-rung is the
singlet dimer state $\phi_{2j}^{(0,0)}$, while that of the ferromagnetic
$J_{\rm rung}$-rung is one of the ferromagnetic states $\phi_{2j-1}^{(1,+)}$
and $\phi_{2j-1}^{(1,-)}$.  The effective interaction between the neighboring
$J_{\rm rung}$-rungs through the in-between $J_{\rm rung}'$-rung can be
obtained by carrying out a third-order perturbation calculation.  The resulting
effective interaction is considered to be antiferromagnetic judging from the
numerical result discussed below (see the right of figure~\ref{fig:mj135}).
Therefore, the $J_{\rm rung}$-rung in the
$\phi_{2j-1}^{(1,+)}$ (or $\phi_{2j-1}^{(1,-)}$) state and that in the
$\phi_{2j+1}^{(1,-)}$ ($\phi_{2j+1}^{(1,+)}$) state arrange
alternatively.  Thus, the ground state in the present case is anticipated to
be the ^^ ferromagnetic'-^^ singlet dimer' (F-SD) state sketched in the right
of figure~\ref{fig:phases}.

\section{Ground-state phase diagram}

We denote, respectively, by $E_0^{(\rm p)}(L,M)$ and $E_1^{(\rm p)}(L,M)$, the
lowest and second-lowest energy eigenvalues of the
Hamiltonian~(\ref{eq:hamiltonian}) under periodic boundary conditions within
the subspace characterized by $L$ and the total magnetization
\hbox{$M\!\equiv\!\sum_{j=1}^{L} (S_{j,a}^z\!+\!S_{j,b}^z)$}.  It is noted that
$E_0^{(\rm p)}(L,0)$ always gives the ground-state energy for the finite-$L$
system.  Then, the excitation energy gap $\Delta_{00}(L)$ within the
{$M\!=\!0$} subspace is given by 
\begin{equation}
  \Delta_{00}(L)=E_1^{(\rm p)}(L,0)-E_0^{(\rm p)}(L,0)\,.
\end{equation}
We also define the site magnetization $m_{j,l}(L)$ as
\begin{equation}
  m_{j,l}(L) = \langle S_{j,l}^z\rangle_L\,,
\end{equation}
where $\langle\cdots\rangle_L$ denotes the ground-state expectation value
for the system with $L$ rungs.

\begin{figure}[t]
\begin{minipage}{23pc}
  \scalebox{0.33}{\includegraphics{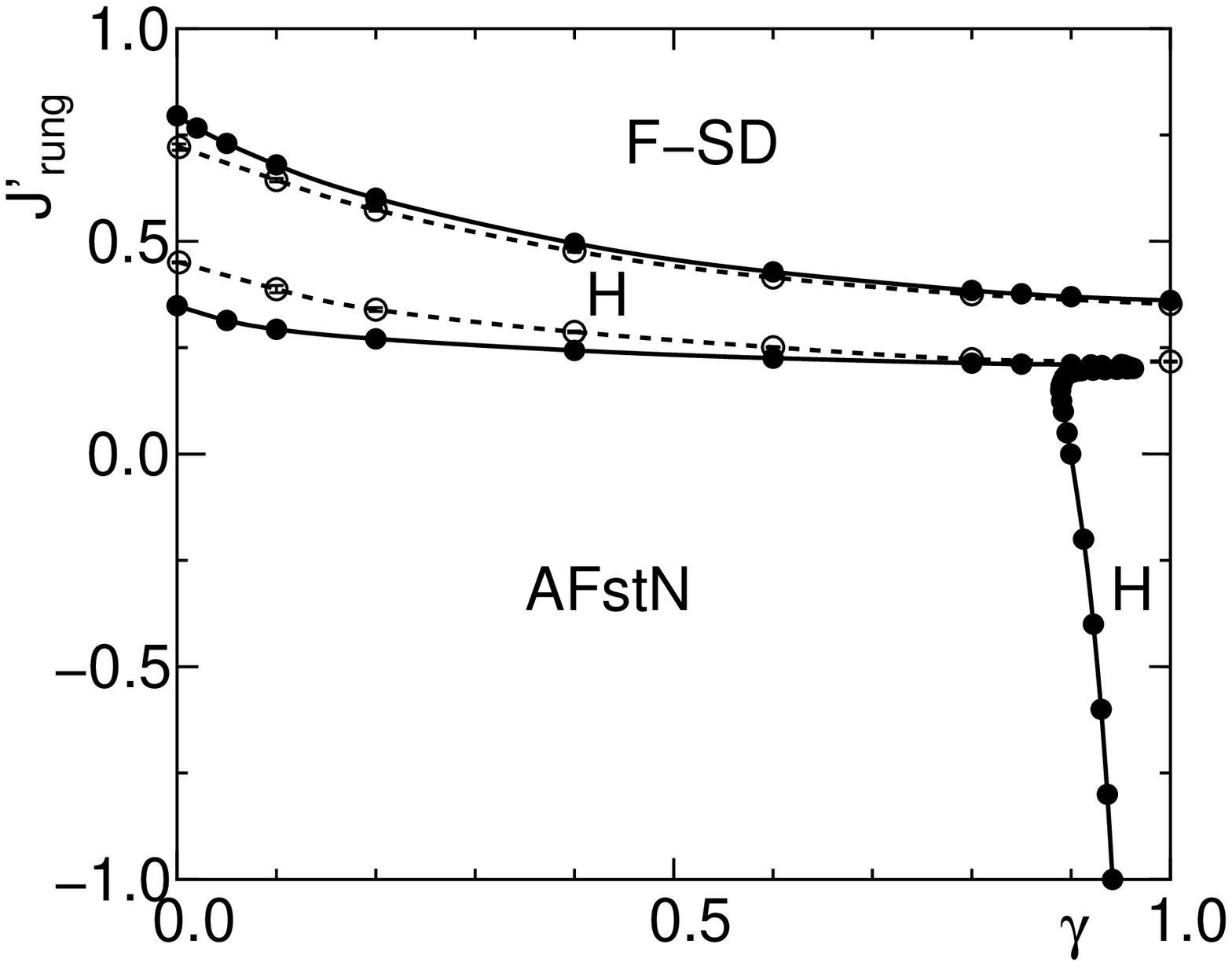}}
  \caption{\label{fig:phase-diagram}
           Ground-state phase diagram on the $\gamma$ versus $J_{\rm rung}'$
           plane.  Here, F-SD, H and AFstN stand, respectively, for
           ^^ ferromagnetic'-^^ singlet dimer', Haldane and antiferromagnetic
           stripe N{\'e}el.
           The solid lines are the second-order (2D Ising) phase boundary
           lines.  The dotted lines, being the Lifshitz lines, separate the
           commensurate and incommensurate regions; the region between these
           lines is the incommensurate one.
           }
\end{minipage}\hspace{2pc}%
\begin{minipage}{11pc}
  \scalebox{0.31}{\includegraphics{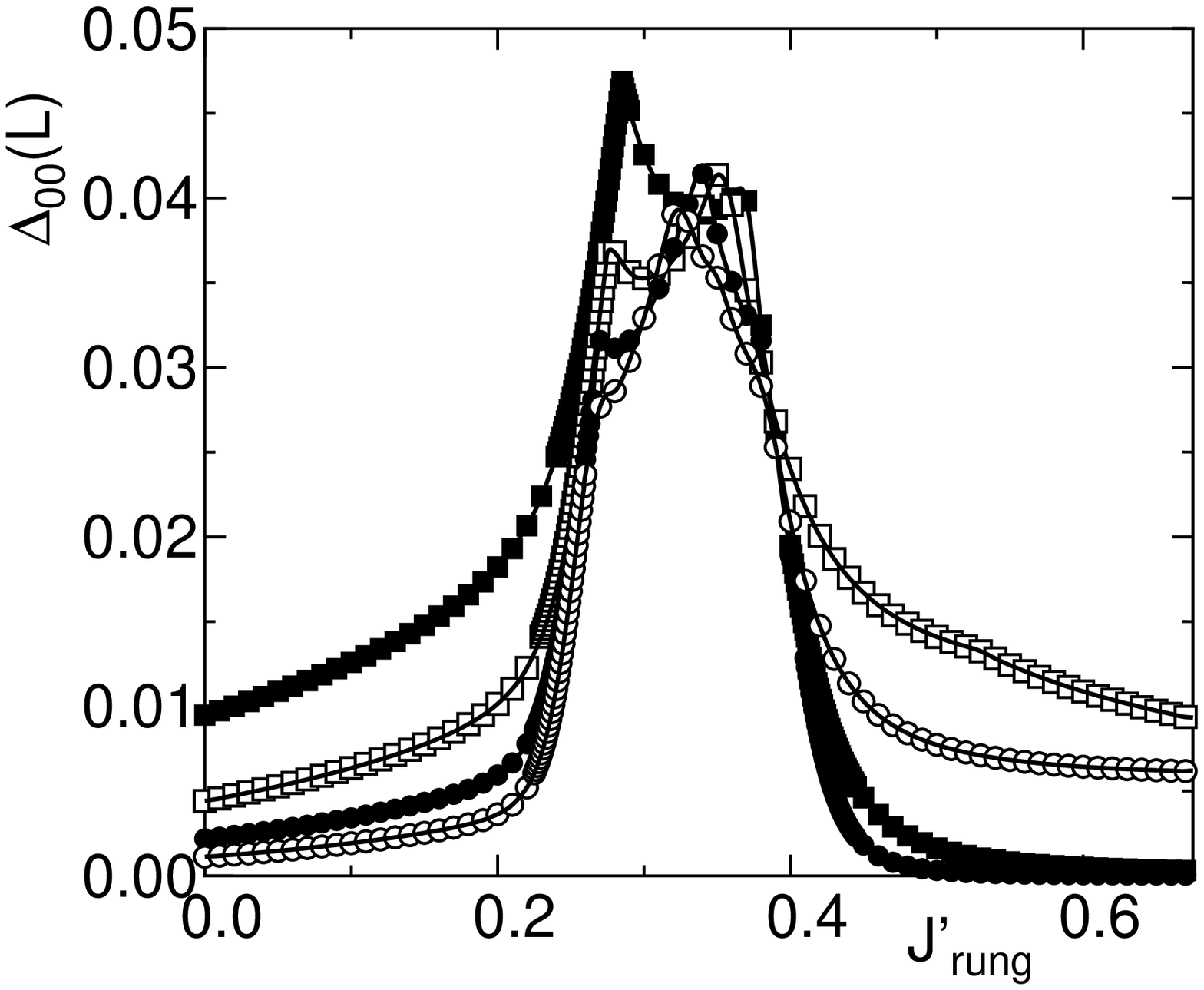}}
  \caption{\label{fig:gap00}
           Plots versus $J_{\rm rung}'$ of the excitation energy gap
           $\Delta_{00}(L)$ for \hbox{$L\!=\!8$} (closed squares), $10$
           (open squares), $12$ (closed circles) and $14$ (open circles)
           along the \hbox{$\gamma\!=\!0.6$} line.
           }
\end{minipage} 
\end{figure}

Figure~\ref{fig:phase-diagram} shows our final result for the
ground-state phase diagram on the $\gamma$ versus $J_{\rm rung}'$
plane.  It consists of the F-SD, Haldane and AFstN phases.  The
solid lines are the 2D Ising phase boundary line between the F-SD and Haldane
phases and that between the Haldane and AFstN phases.  On the other hand, the
dotted lines, which are often called the Lifshitz lines, separate the
commensurate and incommensurate regions; the latter region is between them.
The most striking feature of this phase diagram is the fact that an
incommensurate region appears within the Haldane phase region, which can be
attributed to the frustration effect.  The fact that the Haldane state
appears as the ground state even at the \hbox{$\gamma\!\to\!0$} limit is
also interesting.  This is because the Haldane state is known to become the
ground state mainly in the case of the $XY$-type anisotropy in the
\hbox{$S\!=\!1$} $XXZ$ chain~\cite{chen-etal}.  This phenomenon is called
the inversion phenomenon concerning the interaction
anisotropy~\cite{inv-phenenon-1,inv-phenenon-2,inv-phenenon-3,inv-phenenon-4}.

\begin{figure}[b]
  ~~~~
  \scalebox{0.26}{\includegraphics{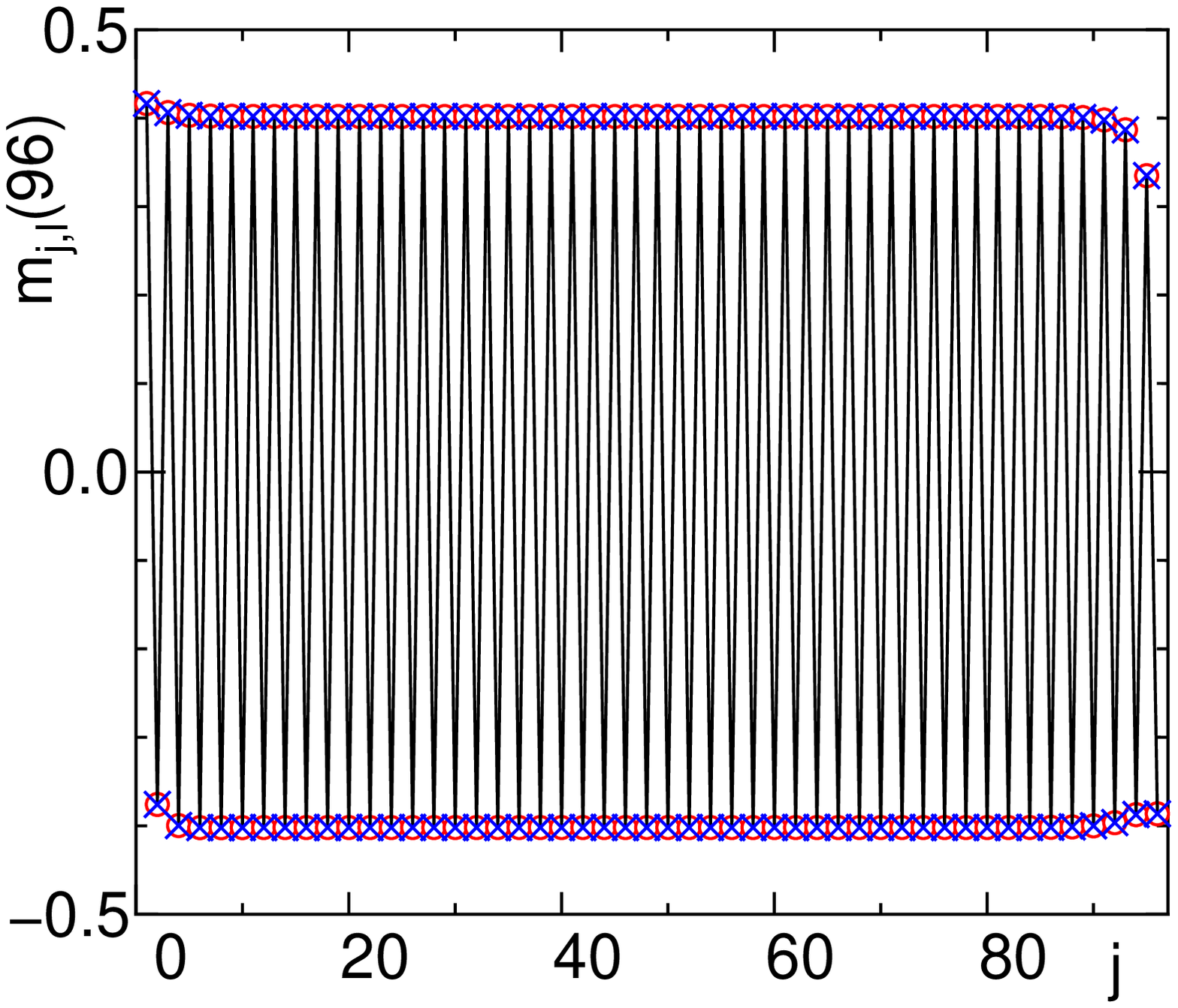}}~~~~~
  \scalebox{0.26}{\includegraphics{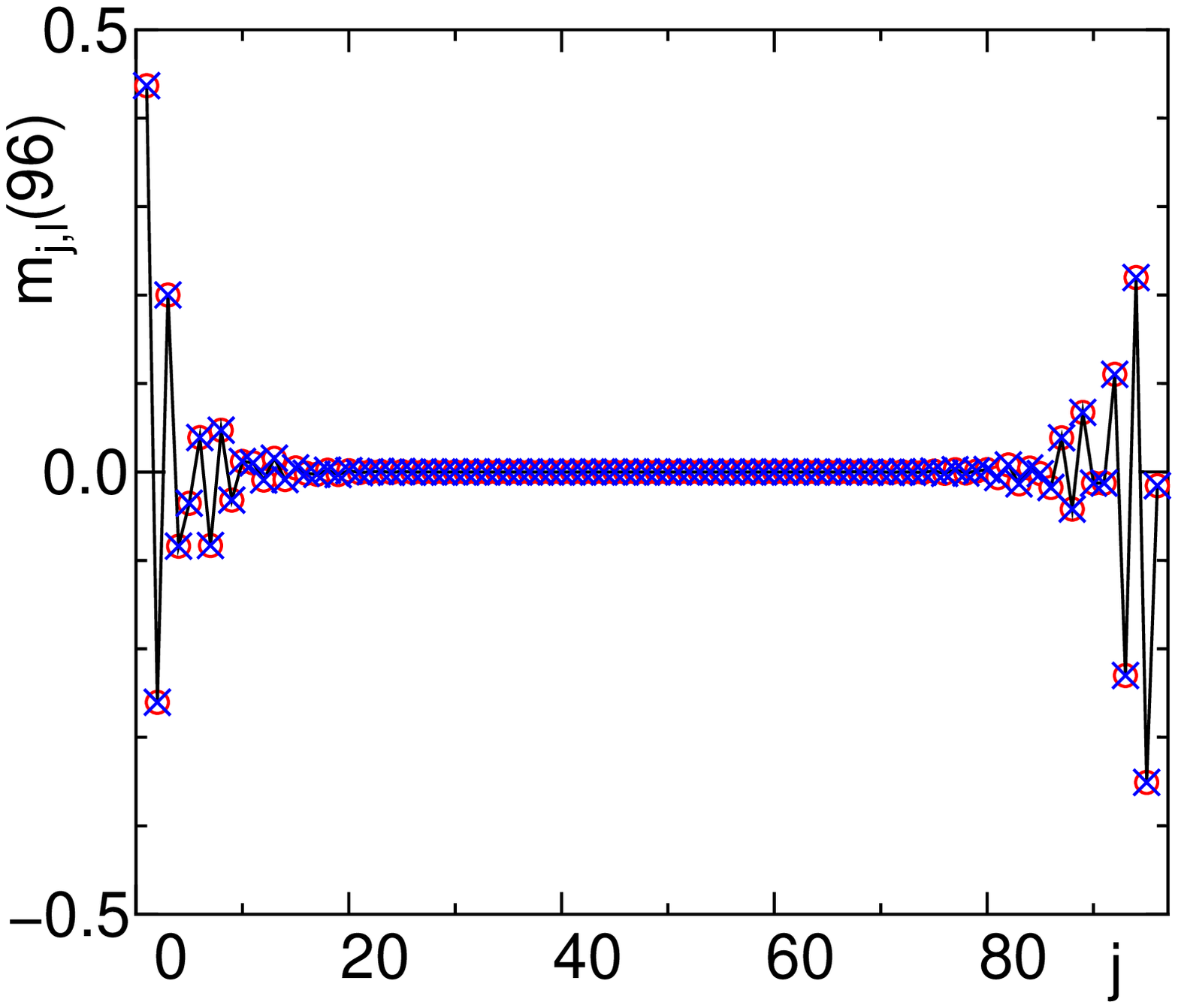}}~~~~~
  \scalebox{0.26}{\includegraphics{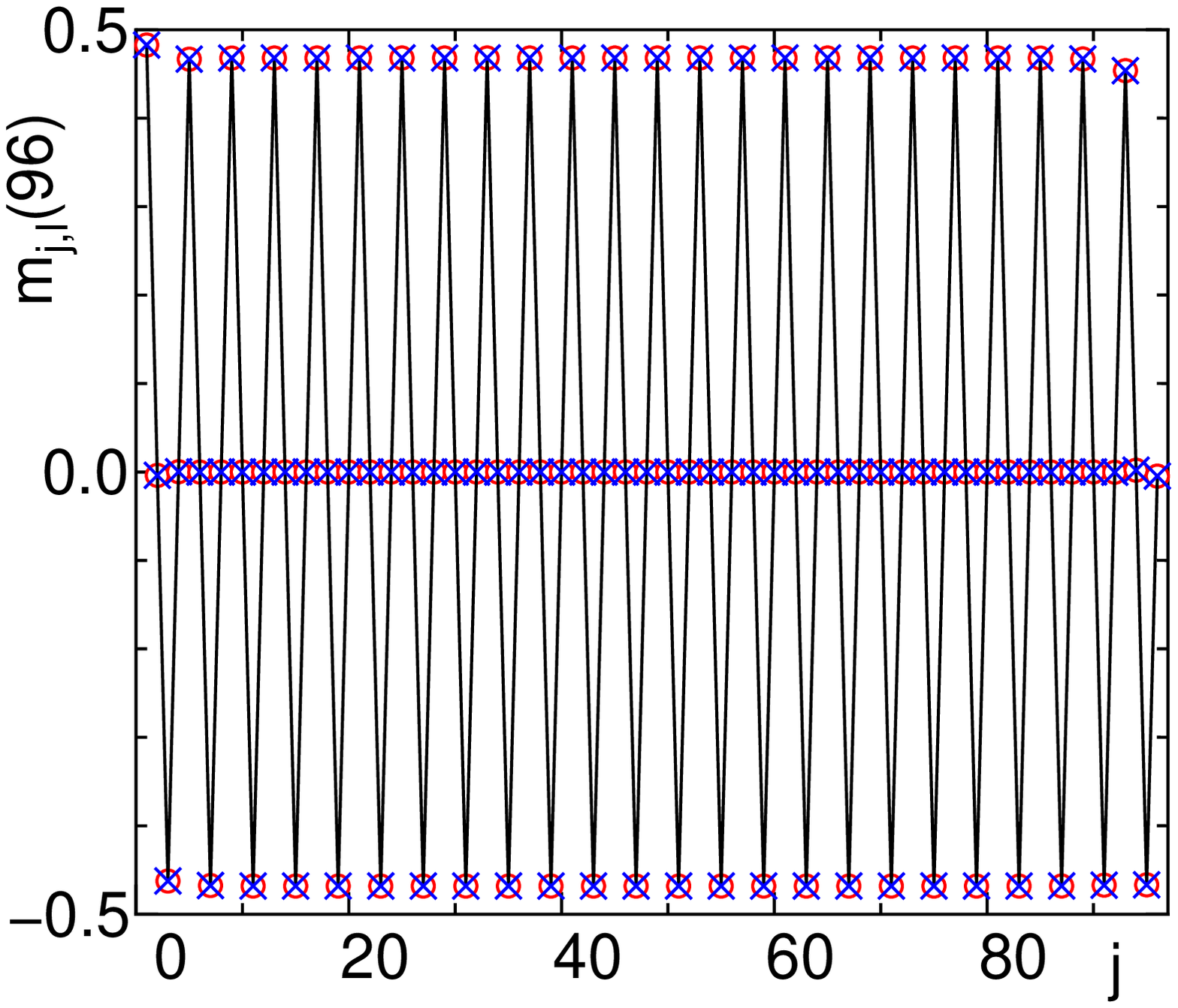}}
  \caption{Plots versus $j$ of the site magnetizations $m_{j,l}(96)$
           (\hbox{$l\!=\!a$}, $b$) for \hbox{$J_{\rm rung}'\!=\!0.1$}
           (left), $0.3$ (middle) and $0.5$ (right) with $\gamma$ fixed
           at \hbox{$\gamma\!=\!0.6$}.  The red circles and the blue crosses
           show, respectively, $m_{j,a}(96)$ and $m_{j,b}(96)$.
           }
\label{fig:mj135}
\end{figure}

Let us turn to a discussion on the determination of the phase diagram.  In
figure~\ref{fig:gap00} we plot the $J_{\rm rung}'$-dependence of the excitation
energy gap $\Delta_{00}(L)$ for \hbox{$\gamma\!=\!0.6$}, calculated by the ED
method.  This figure demonstrates that the ground state is doubly degenerate
when \hbox{$-1\!\leq\!J_{\rm rung}' \lsim 0.22$} and when
\hbox{$0.43 \lsim J_{\rm rung}'\!\le\!1$}, while it is unique when
\hbox{$0.22 \lsim J_{\rm rung}'\lsim 0.43$}.  From these results together with
the physical considerations discussed in section~2, we may expect that, when
\hbox{$-1\!\leq\!J_{\rm rung}' \lsim 0.22$},
\hbox{$0.22 \lsim J_{\rm rung}'\lsim 0.43$} and
\hbox{$0.43 \lsim J_{\rm rung}'\!\le\!1$}, the ground states are, respectively,
the AFstN, Haldane and F-SD states.  In order to ascertain these expectations,
we have performed DMRG calculations for the finite-size system with
\hbox{$2L\!=\!192$} spins under open boundary conditions to evaluate the
site magnetization $m_{j,l}(96)$ for various values of
$J_{\rm rung}'$~\cite{commentDMRG}.  As examples, the results for the cases of
\hbox{$J_{\rm rung}'\!=\!0.1$}, $0.3$ and $0.5$ are depicted in
figure~\ref{fig:mj135}.  We see from this figure that in all cases
$m_{j,a}(96)$ and $m_{j,b}(96)$ are almost equal to each other.  In the first
case, \hbox{$m_{2j-1,a}(96)\!\simeq\!-m_{2j,a}(96)$}, which shows that the
ground state in this case is the AFstN state.  In the second case, the edge
states clearly exist; this is one of the most representative features of the
Haldane state~\cite{edge-states-1,edge-states-2,edge-states-3}.  Finally in
the third case, \hbox{$m_{4j-3,a}(96)\!\simeq\!-m_{4j-1,b}(96)$} and
\hbox{$m_{2j,a}(96)\!\simeq\!0$}, which show that the ground state is the F-SD
state.

Both of the phase transition between the AFstN and Haldane states and that
between the Haldane and F-SD states are of the 2D Ising type, since the $Z_2$
symmetry is broken in the AFstN and F-SD states while it is not broken in the
Haldane state.  It is well known that the phenomenological renormalization
group (PRG) method~\cite{PRmethod} is a useful one to determine the phase
boundary line for this phase transition.  The PRG equations for the
(AFstN,Haldane)-transition and for the
(Haldane,F-SD)-transition are, respectively, given by
\begin{equation}
    L\,\Delta_{00}(L) = (L+2)\,\Delta_{00}(L+2)\,,~~~~~~~~~
    L\,\Delta_{00}(L) = (L+4)\,\Delta_{00}(L+4)\,.
\end{equation}
This is because the periods along each leg are $2$ and $4$ in the AFstN and
F-SD states, respectively.  Solving numerically these PRG equations for
a given value of $\gamma$ (or $J_{\rm rung}'$), we have computed
the finite-size (AFstN,Haldane)-transition point
$J_{\rm rung,c}'^{\rm (AFstN,H)}(L)$
$\bigl(\gamma_{\rm c}^{\rm (AFstN,H)}(L)\bigr)$ for \hbox{$L\!=\!6$}, $8$,
$10$, $12$, and the finite-size (Haldane,F-SD)-transition
point~$J_{\rm rung,c}'^{\rm (H,F-SD)}(L)$
$\bigl(\gamma_{\rm c}^{\rm (H,F-SD)}(L)\bigr)$ for \hbox{$L\!=\!4$}, $8$,
$12$.  We have extrapolated these finite-size data to the thermodynamic
(\hbox{$L\!\to\!\infty$}) limit by fitting them for the former and for the
latter to quadratic functions of $(L\!+\!1)^{-2}$ and of $(L\!+\!2)^{-2}$,
respectively.   Some examples of the results are
\hbox{$J_{\rm rung,c}'^{\rm (AFstN,H)}(\infty)\!=\!0.225\pm 0.001$} and
\hbox{$J_{\rm rung,c}'^{\rm (H,F-SD)}(\infty)\!=\!0.428\pm 0.001$} for
\hbox{$\gamma\!=\!0.6$}, and also
\hbox{$\gamma_{\rm c}^{\rm (AFstN,H)}(\infty)\!=\!0.942\pm 0.002$} for
\hbox{$J_{\rm rung}'\!=\!-1.0$}.  It is noted that the above value of
$\gamma_{\rm c}^{\rm (AFstN,H)}(\infty)$ for
\hbox{$J_{\rm rung}'\!=\!-1.0$} is in fairly good agreement with the
corresponding value $\sim\!0.9$ obtained by mapping the present
Hamiltonian~(\ref{eq:hamiltonian}) onto the effective
Hamiltonian~(\ref{eq:effective}) for the \hbox{$S\!=\!1$} chain (see
subsection~2.1).  We have carried out the same procedure for various values of
$\gamma$ and $J_{\rm rung}'$ to obtain the 2D Ising phase boundary lines shown
by the solid lines in figure~\ref{fig:phase-diagram}.

\begin{figure}[b]
\begin{minipage}{21pc}
  \scalebox{0.26}{\includegraphics{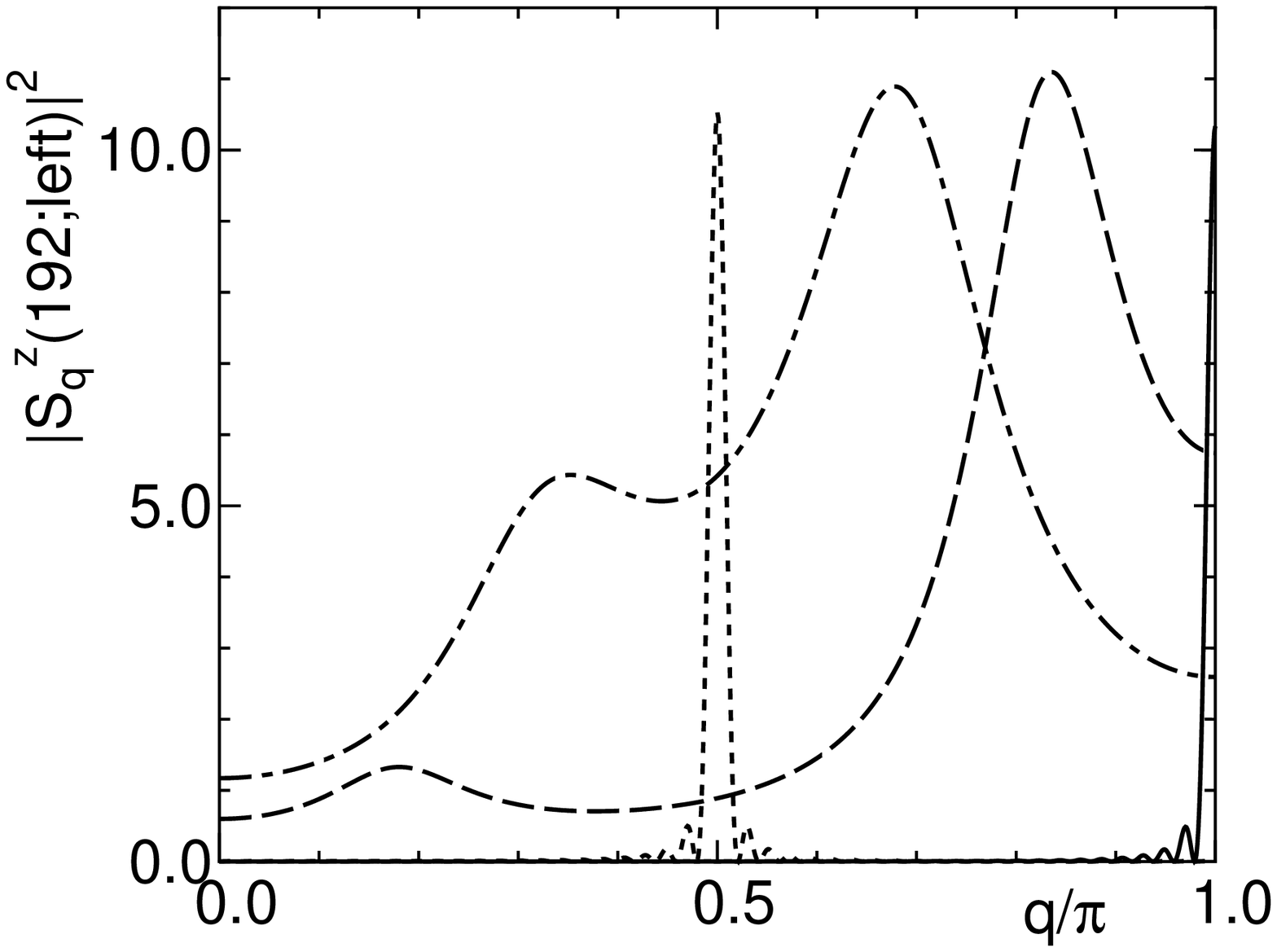}}
  \caption{\label{fig:sq-curve}
           Squared modulus of the Fourier transform of the site magnetization,
           $\vert{\cal S}_q^z(192;{\rm left})\vert^2$, plotted versus $q/\pi$
           with $\gamma$ fixed at
           \hbox{$\gamma\!=\!0.6$}.  The dotted, dot-dashed, dashed and solid
           lines show, respectively, the results for
           \hbox{$J_{\rm rung}'\!=\!0.1$} (divided by 3), those for
           \hbox{$J_{\rm rung}'\!=\!0.3$} (multiplied by 460), those for
           \hbox{$J_{\rm rung}'\!=\!0.37$} (multiplied by 900) and those for
           \hbox{$J_{\rm rung}'\!=\!0.5$}.
           }
\end{minipage}\hspace{2pc}%
\begin{minipage}{12pc}
  \scalebox{0.26}{\includegraphics{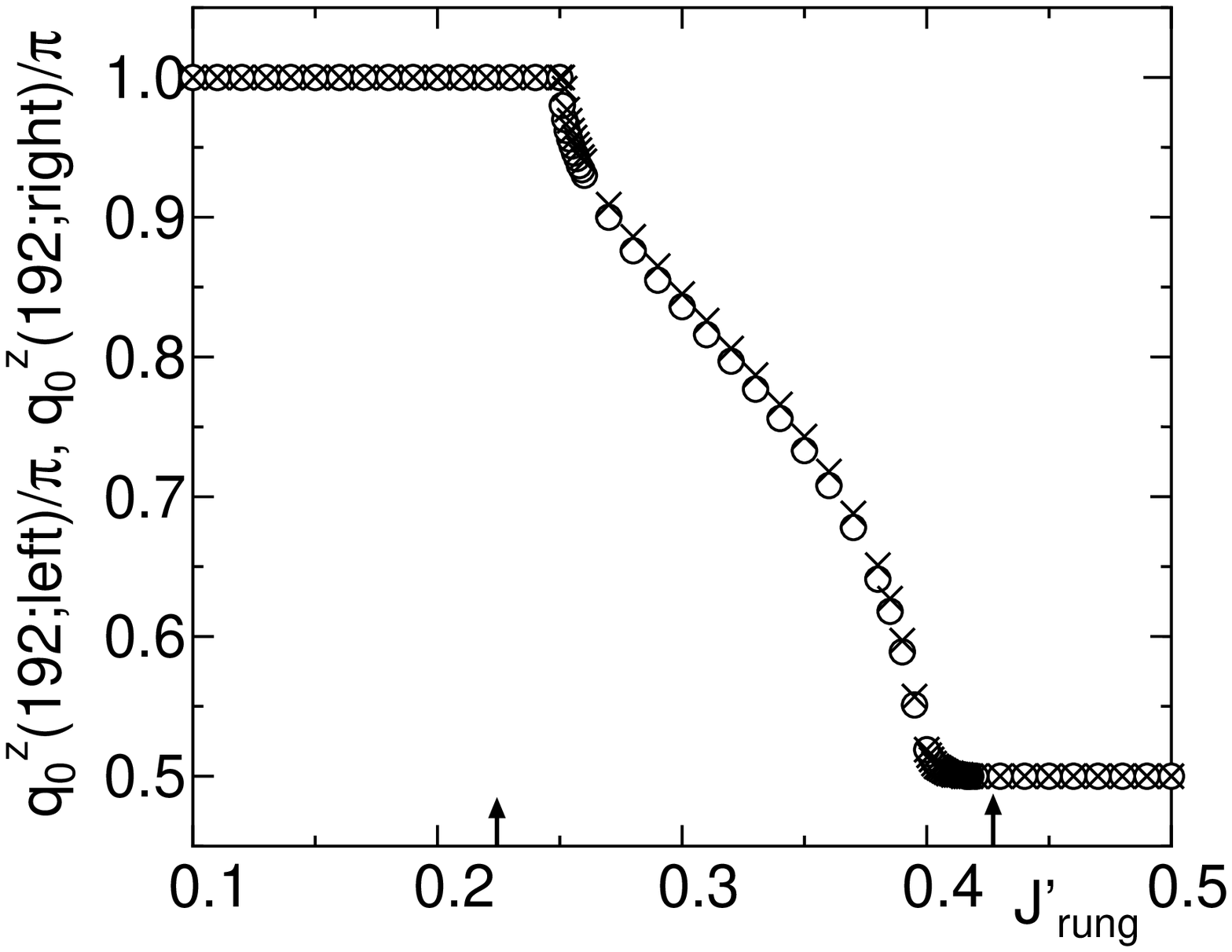}}
  \caption{\label{fig:q0-curve}
           Plots versus $J_{\rm rung}'$ of $q_0^z(192;{\rm left})$
           (circles) and $q_0^z(192;{\rm right})$ (crosses) with
           $\gamma$ fixed at \hbox{$\gamma\!=\!0.6$}.  The two arrows on the
           abscissa indicate the (AFstN,Haldane)-transition and
           (Haldane,F-SD)-transition points.
           }
\end{minipage} 
\end{figure}

The Lifshitz line which separates the commensurate and incommensurate regions
can be estimated by examining the Fourier transform of the site magnetization
$m_{j,l}(L)$~\cite{hikihara-etal-1,hikihara-etal-2}.  When we adopt open
boundary conditions, the present \hbox{$S\!=\!1/2$} rung-alternating ladder
system has no inversion symmetry with respect to its center position, that is,
\hbox{$m_{j,l}(L)\!\ne\!m_{L+1-j,l}(L)$}.  Therefore, it is important,
especially in the region of the Haldane phase, to discuss two kinds of the
Fourier transforms, ${\cal S}_q^z(L;{\rm left})$ and
${\cal S}_q^z(L;{\rm right})$, defined by
\begin{equation}
   {\cal S}_q^z(L;{\rm left}) = {1\over \sqrt{L/2}}
             \sum\nolimits_{j=1}^{L/2} \exp(iqj)\, m_j(L)\,,~~~
   {\cal S}_q^z(L;{\rm right}) = {1\over \sqrt{L/2}}
             \sum\nolimits_{j=1}^{L/2} \exp(iqj)\, m_{L+1-j}(L)\,\,\,
\end{equation}
with \hbox{$m_j(L)\!=\!m_{j,a}(L)\!+\!m_{j,b}(L)$}, in order to avoid the
mismatch of the edge states of both edges.  In the above equations $q$ is
the wave number.

In figure~\ref{fig:sq-curve} the $q$-dependences of the squared modulus of
${\cal S}_q^z(192;{\rm left})$, which have been calculated by using the DMRG
results for $m_{j,l}(192)$ for the finite-size system with
\hbox{$2L\!=\!384$} spins, are shown for the cases of
\hbox{$J_{\rm rung}'\!=\!0.1$}, $0.3$, $0.37$ and $0.5$.  It is noted that
the $q$-dependences of ${\cal S}_q^z(192;{\rm right})$ show almost the same
behavior.  Thus, the values $q_0^z(192;{\rm left})$ and
$q_0^z(192;{\rm right})$ of $q$ which give, respectively, the maximum values
of $\vert{\cal S}_q^z(192;{\rm left})\vert^2$ and
$\vert{\cal S}_q^z(192;{\rm right})\vert^2$ are almost equal to each other.
Figure~\ref{fig:q0-curve} depicts the plots versus $J_{\rm rung}'$ of
$q_0^z(192;{\rm left})$ and $q_0^z(192;{\rm right})$ for
$\hbox{$\gamma\!=\!0.6$}$.  From this figure we clearly see that, there are two
Lifshitz points $J_{\rm rung,Lifshitz}'^{(1)}(192)$ and 
$J_{\rm rung,Lifshitz}'^{(2)}(192)$, and in the region of
\hbox{$J_{\rm rung,Lifshitz}'^{(1)}(192)\!\leq\!J_{\rm rung}'\!\leq\!
J_{\rm rung,Lifshitz}'^{(2)}(192)$}, the system has the incommensurate
character in the sense that both of $q_0^z(192;{\rm left})$ and
$q_0^z(192;{\rm right})$ are larger than $\pi/2$ and smaller than $\pi$, where
\hbox{$J_{\rm rung,Lifshitz}'^{(1)}(192)\!=\!0.251\pm 0.001$} and
\hbox{$J_{\rm rung,Lifshitz}'^{(2)}(192)\!=\!0.415\pm 0.001$}.  These values
of the Lifshitz points yield good approximations for the corresponding
\hbox{$L\!\to\!\infty$} ones, since our calculations show that the finite-size
values for \hbox{$L\!=\!192$}, $96$ and $72$ agree with each other within
numerical errors.  We have performed the same procedure for various
values of $\gamma$, and obtained the Lifshitz lines shown by the
dotted lines in figure~\ref{fig:phase-diagram}.  It is noted that,
in the region between these two Lifshitz lines, the ground state of the present
system has an incommensurate character.

\section{Summary}

We have numerically determined, with the help of some physical considerations,
the ground-state phase diagram of an anisotropic rung-alternating
\hbox{$S\!=\!1/2$} ladder, which is described by the
Hamiltonian~(\ref{eq:hamiltonian}), in the case where
\hbox{$J_{\rm leg}\!=\!0.2$}, \hbox{$J_{\rm rung}\!=\!-1$},
\hbox{$|J_{\rm rung}'|\!\leq\!1$} and \hbox{$0\!\leq\!\gamma\!<\!1$}.
The obtained phase diagram on the $\gamma$ versus $J_{\rm rung}'$ plane is
shown in figure~\ref{fig:phase-diagram}.

\ack

We would like to express our sincere thanks to Professors U Schollw{\"o}ck,
T Vekua, K Hida and J Richter for their invaluable discussions and comments.
We also thank the Supercomputer Center, Institute for Solid State Physics,
University of Tokyo and the Computer Room, Yukawa Institute for Theoretical
Physics, Kyoto University for computational facilities.  Finally,
T.H.~acknowledges the support by JSPS KAKENHI Grant Nunmber 15K05198.


\begin{thebibliography}{9}
\bibitem{commentDMRG}
We note that, fundamentally speaking, the site magnetization $m_{j,l}(L)$
must be zero in the ground state of the finite-$L$ system.  However, when the
ground states are degenerate (or the energy gap is exponentially small), the
DMRG calculation for a finite system often selects one of them: In our
calculation for the $M=0$ subspace, we have obtained one of the magnetic
ordered states in the F-SD and AFstN phases and one of the ground states in
which free effective spins at open edges are polarized in opposite directions
in the Haldane phase.
\end{thebibliography}
\end{document}